\documentclass[pre,aps,showpacs,superscriptaddress,preprint]{revtex4}
\usepackage{graphicx,color}

\begin{document}
\definecolor{red}{rgb}{1,0,0}
\definecolor{green}{rgb}{0,1,0}
\title{Quantum versus Classical Dynamics in a driven barrier: the role of kinematic effects}

\author{P.K. Papachristou}
\affiliation{Department of Physics, University of Athens, GR--15771,
Athens,
Greece}

\author{E. Katifori}
\affiliation{Center for Studies in Physics and Biology, The Rockefeller University, 
New York, New York 10065, USA}

\author{F.K. Diakonos}
\affiliation{Department of Physics, University of Athens, GR--15771,
Athens,
Greece}

\author{V. Constantoudis}
\affiliation{Institute of Microelectronics (IMEL), NCSR "Demokritos",
P.O. Box 60228,
Aghia Paraskevi, Attiki, Greece 15310}

\author{E. Mavrommatis}
\affiliation{Department of Physics, University of Athens, GR--15771,
Athens,
Greece}

\begin{abstract}
We study the dynamics of the classical and quantum mechanical scattering of a wave packet from an oscillating barrier. Our main focus is on the dependence of the transmission coefficient on the initial energy of the wave packet for a wide range of oscillation frequencies. The behavior of the quantum transmission coefficient is affected by tunneling phenomena, resonances and kinematic effects emanating from the time dependence of the potential. We show that when kinematic effects dominate (mainly in intermediate frequencies), classical mechanics provides very good approximation of quantum results. In that frequency region, the classical and quantum transmission coefficients are in optimal agreement. Moreover, the transmission threshold, i.e. the energy above which the transmission coefficient becomes larger than a specific small threshold value, is found to exhibit a minimum. We also consider the form of the transmitted wave packet and we find that for low values of the frequency the incoming classical and quantum wave packet can be split into a train of well separated coherent pulses, a phenomenon which admits purely classical kinematic interpretation.
\end{abstract}
\maketitle

\section{Introduction}

The agreement between quantum and classical mechanics, usually referred to as Quantum-Classical Correspondence (QCC) is a subject that has intensively been studied in a large variety of systems. In the semiclassical limit, classical mechanics is a useful tool for the study of quantum systems, especially in the case where ab initio quantum calculations are lengthy and cumbersome \cite{GRO96,FAG08}. It is therefore important to know under which circumstances a quantum system may be adequately described by classical mechanics. The majority of works on QCC have been devoted to bound or semi-bound systems with static potentials, and less attention has been devoted to scattering systems \cite{LEW04,BAC05,LUN02} as well as to systems with time-dependent potentials \cite{HOL95,SHA99}. The behaviour of such systems cannot be fully attributed to the shape of the potential, as kinematic effects may play an important role \cite{YAN04,SPA09}.

The aim of the present work is to study QCC in a simple time-dependent system with the presence of kinematic effects. In particular, we study the scattering of classical and quantum wave packets off a 1-dimensional barrier whose position oscillates laterally and harmonically with time. The problem of a charged particle interacting with a static potential barrier in the presence of an oscillating electric field, can be transformed to that of a particle interacting with an oscillating potential barrier, by means of the Kramers-Henneberger transformation \cite{HEN58,GAV92,HEN01}. The interest in the behavior of driven barrier systems has been renewed due to the effect of quantum charge pumping, according to which, in systems of mesoscopic scale subject to an ac driving, a dc current can be generated even at zero bias \cite{BRO98,TOR05,MOS08,BYR12}.

A commonly used approach for the quantification of QCC is the construction of a phase space representation of quantum mechanics and the comparison of the evolution of classical and quantum densities in phase space \cite{WIG32,HUS40,BAL90,BAC04}. Such representations, based on the Wigner and Husimi phase space densities, elucidate the effects of classical phase space on the quantum evolution. However, in most cases of practical interest, the agreement between classical and quantum mechanics is considered with respect to specific observables, such as ionization or dissociation rates \cite{SIR02,CON01,LIM08,FU02}, tunneling probabilities \cite{PAL07}, dwell times \cite{LEW04} etc.

In the case of scattering from barriers, the most widely used observable in both classical and quantum approaches is the transmission coefficient. For a static barrier, the classical transmission coefficient as a function of the energy $E$ of the particles exhibits the form of a step function: it is zero for $E<V_0$ and unity for $E>V_0$, where $V_0$ is the height of the barrier. Nevertheless, if classical wave packets (i.e. ensembles of orbits) are considered, the transmission coefficient can become a continuous function of the mean energy, as it is the case in quantum mechanics. This occurs when the classical wave packet is broad enough in momentum space to include orbits with energies larger than the height of the barrier. Introducing time-dependence, by means of the lateral oscillation of the barrier, will in general enhance the transmission since particles with $E<V_0$ can be transmitted if the energy corresponding to their relative motion with respect to the barrier is greater than $V_0$.

The transmission of wave packets through a laterally harmonically oscillating barrier exhibits very interesting properties and has been studied in several works, mainly in the framework of quantum mechanics \cite{PIM91,GE96,CHI03,VOR98,PAP05,YEA00,GAR10}. It has been found that for high driving frequencies, the transmission coefficient exhibits peaks at energies well below the barrier height. These peaks correspond to resonances which are associated with quasistable bound states of the effective time-averaged potential \cite{CHI03,VOR98}. For intermediate frequencies, inelastic processes dominate the scattering dynamics and strong sidebands appear in the energy spectrum \cite{CHI03}. Recently, the classical mechanics of the system has been extensively studied \cite{PAP05,KOC08}. In particular, it has been found that at high driving frequencies, the system exhibits dynamical trapping which is associated with the existence of a stable island in phase space. As a consequence, the system exhibits non-hyperbolic chaotic scattering as well as stickiness of scattering trajectories in the vicinity of the stable island. The transmission of wave packets through an oscillating barrier has been studied both in the context of classical and quantum mechanics. In such systems, the oscillation frequency introduces an additional time scale and its influence on the QCC is an open question which we attempt to address in the present work. Moreover, it is interesting to investigate whether classical characteristics other than phase space structures have an influence in the quantum mechanics. Such characteristics, as kinematic effects, are investigated in the present work and are shown to play an important role in the dynamics of a time-dependent system in the absence of phase space structures. 

More specifically, in the present work, we compare classical and quantum dynamics for several values of the driving frequency, focusing mainly on the effects that are induced by kinematics rather than by the underlying classical phase space. Our study is centered on the following points:
\begin{enumerate}
\item We compare the classical and quantum transmission coefficient as a function of the incoming wave packet energy for several values of the driving frequency. We find that classical and quantum mechanics exhibit a good quantitative agreement in an intermediate range of frequencies where kinematic effects dominate. Moreover, in this frequency range, the relatively small differences between classical and quantum mechanics can be interpreted, at least at a qualitative level, using mainly classical kinematic arguments. On the contrary, at low and high frequencies classical and quantum behaviors deviate due to the presence of quantum phenomena such as tunneling and resonances.

\item We study the form of the transmitted part of the wave packet as a function of the driving frequency. It is found that in a region of the parameter space, the incoming wave packet can be split to form a train of distinct coherent pulses. This phenomenon appears in both classical and quantum mechanics and we show that it admits a purely classical interpretation. The dependence of this phenomenon on the parameters of the system as well as its possible applications and experimental realizations are also investigated.

\end{enumerate}
Both phenomena investigated in this work (QCC in the transmission coefficient and in the formation of coherent pulse trains) can be interpreted using classical kinematic arguments rather than phase space effects.

The remainder of the present paper is organized as follows. In Sec. 2 we describe our model system and the methodology of our study. In Sec. 3 we present the study of the classical and quantum transmission coefficients. In Sec. 4 we describe the formation of coherent pulse trains. Finally, in Sec. 5 we summarize our findings.

\section{Description of the system and methodology}
\begin{figure}
\begin{center}
\includegraphics[width=9cm]{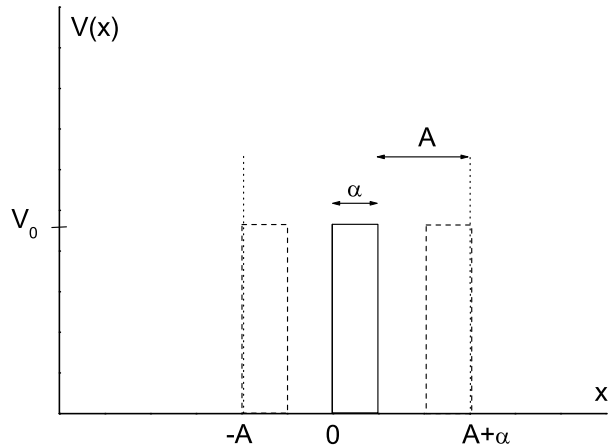}
\caption{The oscillating barrier at its equilibrium position (solid line) and its extremal positions (dashed line). The boundaries of the interaction region are shown with the dotted line.\label{bar_desc}}
\end{center}
\end{figure}

Our system consists of a particle of mass $m$ interacting with a laterally oscillating repulsive potential barrier. The oscillation of the barrier is harmonic with amplitude $A$ and the system is described by the Hamiltonian:
\begin{equation}
H(x,p,t) = \frac{{p^2 }}{{2m}} + V\left( {x - A\sin \left( {\omega t} \right)} \right), \label{ham}
\end{equation}
where $V(x)$ has been chosen to have the form of a rectangular barrier of height $V_0$ and width equal to $\alpha$ (see Fig.~\ref{bar_desc}):
\begin{equation}
V(x) = \left\{ {\begin{array}{*{20}c}
   {V_0 ,{\rm   0 < x < }\alpha~~~~~~~~{\rm          }}  \\
   {{\rm 0,   x} \le {\rm 0 }~~{\rm and}~~{\rm  x} \ge \alpha }  \\
\end{array}} \right. .
\end{equation}
The problem of a charged particle interacting with a static potential barrier in the presence of a spatially uniform alternating electric field, as mentioned above, can be transformed, by means of the Kramers-Henneberger transformation \cite{HEN58,GAV92,HEN01}, to that described by the Hamiltonian (\ref{ham}), with
\begin{equation}
A = \frac{{q\mathcal{E}}}{{m\omega ^2 }},
\end{equation}
where $q$ is the charge of the particle, $\mathcal{E}$ is the amplitude of the electric field and $\omega$ is the frequency of the field oscillation.

We will mainly use the following values of the parameters: $A=200$, $\alpha=80$, $V_0=0.0147$ and $m=0.1$. All quantities are given in atomic units. These values of the parameters are adapted to the conditions of electron transmission from a AlGaAs-GaAs structure in the presence of a laser field \cite{TSU73,CHA74}. We study the transmission of classical and quantum wave packets from the oscillating potential barrier. For the study of classical wave packets, we use as initial conditions ensembles of orbits having positions and momenta that follow Gaussian probability distributions. These initial conditions are integrated forward in time by solving Hamilton's equations. For the study of the quantum wave packets, we use Gaussian wave packets and we solve the time-dependent Schr\"odinger equation using the Crank-Nicolson finite difference scheme combined with mask functions in order to avoid artificial reflections of the wave packets at the boundaries of our spatial grid \cite{KOO90,KRA92,ERM99}. 

\section{Classical and quantum transmission coefficient}
In this section we discuss our calculations of the classical and quantum mechanical transmission coefficient. At the quantum level, our initial state is a gaussian wave packet which is broad in position space ($\sigma_x=5000$) and, as a consequence, narrow in momentum space. Its center at $t=0$ is located at the position $x_0=-A-3\sigma_x$, i.e. the right tail of the wave packet is at the left boundary of the interaction region, which is defined as the region in which the particles can interact with the oscillating potential (see Fig.~\ref{bar_desc}). At the classical level, we evolve in time an ensemble of initial conditions having the same probability density in phase space with the quantum wave packet:
\begin{equation}
\rho (x,p,t = 0) = \frac{1}{{\sqrt {2\pi } \sigma _x }}{\mathop{\rm e}\nolimits} ^{ - \frac{{(x - x_0 )^2 }}{{2\sigma _x ^2 }}} \frac{1}{{\sqrt {2\pi } \sigma _p }}{\mathop{\rm e}\nolimits} ^{ - \frac{{(p - p_0 )^2 }}{{2\sigma _p ^2 }}},
\end{equation}
where $p_0$ is the initial mean momentum of the wave packet. We use minimal uncertainty wave packets, i.e. $\sigma _x \sigma _p  = {\textstyle{1 \over 2}}$. 

\begin{figure}
\begin{center}
\includegraphics[width=7.0cm]{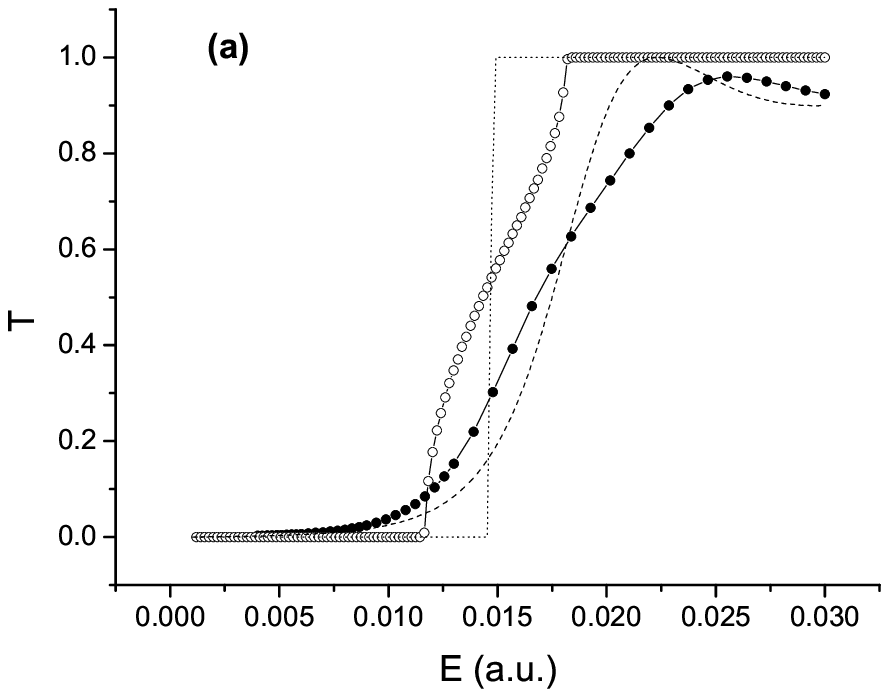}
\includegraphics[width=7.0cm]{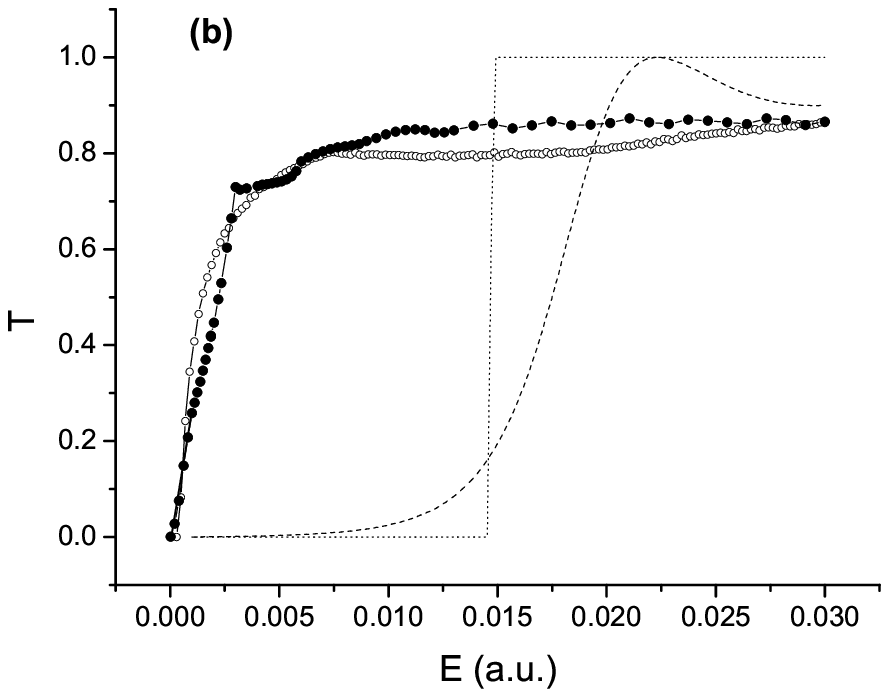}
\includegraphics[width=7.0cm]{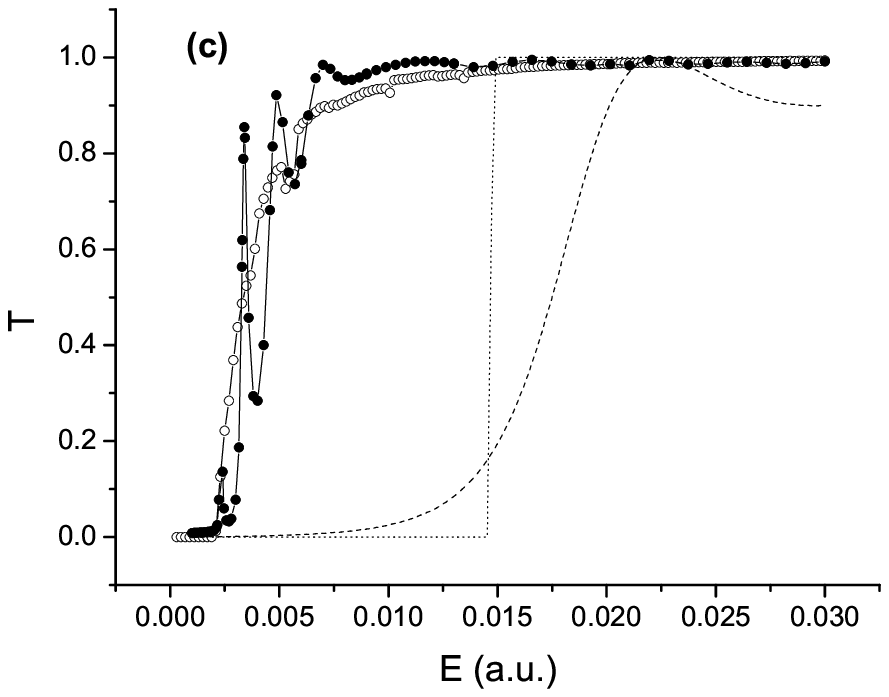}
\caption{\label{bar_tce} Classical (hollow circles) and quantum (full circles) transmission coefficient as a function of the energy of the incoming wave packet for three frequency values, namely (a) $\omega=3\cdot 10^{-4}$, (b) $\omega=3\cdot 10^{-3}$ and (c) $\omega=3\cdot 10^{-2}$. The classical (dotted line) and quantum (dashed line) transmission coefficients for the static barrier are also shown.}  
\end{center}
\end{figure}

We define a driving frequency $\omega_I$ as
\begin{equation}
\omega _I  = \sqrt {\frac{{2V_0 }}{{mA^2 }}} .
\end{equation}
For $\omega=\omega_I$, the barrier can be penetrated even from a particle at rest, colliding with the barrier when the phase of its oscillation is equal to $\pi$. In our system, for the values of the parameters chosen, $\omega_I\simeq 2.7\cdot 10^{-3}$. In the following, we will refer to the frequency range where $\omega\simeq\omega_I$ as intermediate frequency range, whereas frequencies for which $\omega\ll \omega_I$ and $\omega\gg \omega_I$, will be referred to as low and high frequencies respectively. Initially, we present the results for the transmission coefficient for three values of the oscillation frequency, namely $\omega=3\cdot 10^{-4}$, $\omega=3\cdot 10^{-3}$ and $\omega=3\cdot 10^{-2}$, as a function of the mean initial energy of the wave packet. These values of the frequency are in the low, intermediate and high frequency ranges as have been previously defined. Our results are shown in Fig.~\ref{bar_tce}, along with those for the static barrier.

In the case of the classical wave packet interacting with the static barrier, the transmission coefficient is a smooth function of the energy (see the dotted line in Fig.~\ref{bar_tce}(a)). As discussed before, this occurs because a wave packet, even if its expectation value of the energy is smaller than the barrier height $V_0$, can include orbits with energy larger than $V_0$. Assuming a gaussian distribution for the momenta, the transmission coefficient of a wave packet with mean momentum $P$ is given by 
\begin{equation}
T(P) = \frac{1}{{\sqrt {2\pi } \sigma _p }}\int\limits_{\sqrt {2mV_0 } }^\infty  {\exp \left[ { - \frac{{\left( {p - P} \right)^2}}{{2\sigma _p^2 }}} \right]} dp,
\end{equation}
where the mean momentum $P$ and mean energy $E$ are related by
\begin{equation}
E = \frac{{P^2  + \sigma _p^2 }}{{2m}}.
\end{equation}
In our case, since the wave packet is very narrow in momentum space, the classical transmission coefficient increases rapidly for $E\simeq V_0$ and the result shown in Fig.~\ref{bar_tce}(a) (dotted line) is close to a step function. The corresponding quantum curve (dashed line) is smoother due to quantum tunneling and interference.

From Fig.~\ref{bar_tce} we observe that, in general, the increase of the driving frequency leads to an overall enhancement of transmission, due to the increase of the mean energy of the motion of the particles with respect to the barrier. Nevertheless, apart from this general trend, the dependence of the transmission coefficient on the frequency and on the incident energy is more complicated. Interestingly, we observe that the agreement between classical and quantum mechanics is better in the case of the intermediate frequencies (see Fig.~\ref{bar_tce}(b)). In that case there is very good qualitative agreement almost in the whole range of the energies considered and a relatively good quantitative agreement: the sum of the squares of the quantum-classical differences of the transmission coefficient is $0.49$ whereas for the low and high frequency the corresponding values are $4.16$ and $1.15$ respectively. At low frequencies (see Fig.~\ref{bar_tce}(a)), for energies below the onset of classical transmission, tunneling leads to a non-vanishing quantum transmission coefficient, whereas for larger energies transmission is suppressed due to quantum interference. In the case of high frequencies (see Fig.~\ref{bar_tce}(c)), the quantum transmission coefficient exhibits four peaks that are not apparent in the corresponding classical calculation. Such peaks have been reported and explained in \cite{VOR98}. The explanation is based on the fact that at the limit of high frequencies, the scattered particle feels an effective static potential which is the time average of the oscillating potential and in our case it has the form of a double barrier. The peaks correspond to resonances in the double barrier. 

\begin{figure}
\begin{center}
\includegraphics[width=12cm]{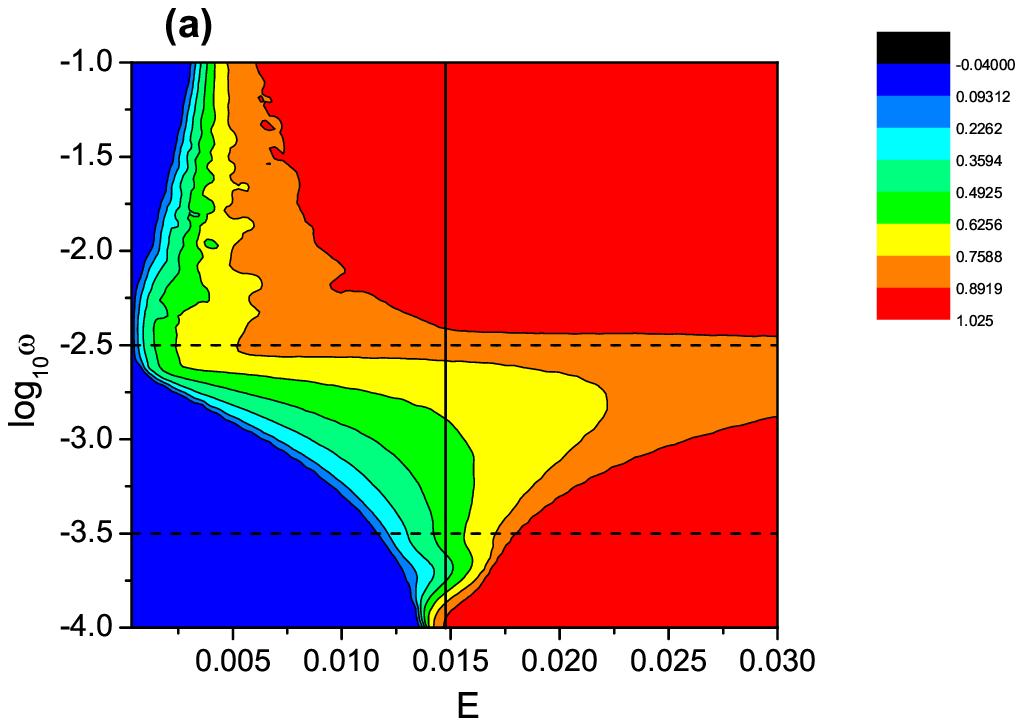}
\includegraphics[width=12cm]{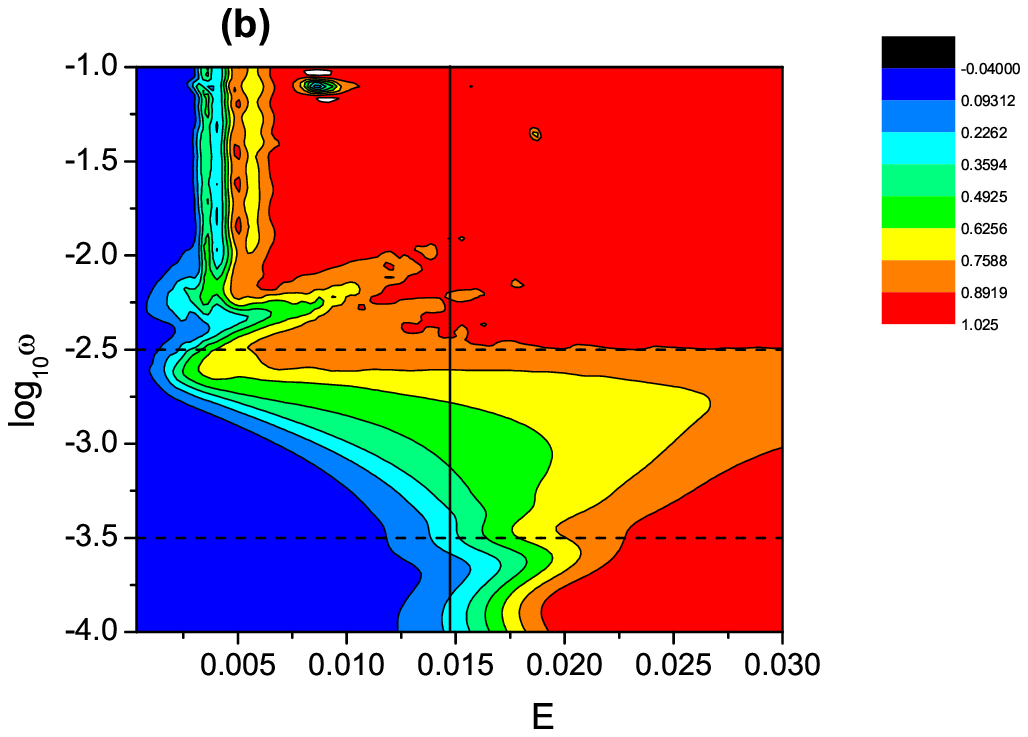}
\caption{\label{tc2D} Contour plots of the (a) classical and (b) quantum transmission coefficients as a function of the energy $E$ of the incoming wave packet and of the frequency of the oscillation $\omega$. The height of the potential barrier is shown as a solid vertical line. The horizontal dashed lines correspond to frequencies $\log_{10}\omega=-2.5$ and $\log_{10}\omega=-3.5$ (see text).}
\end{center}
\end{figure}

\begin{figure}
\begin{center}
\includegraphics[width=12cm]{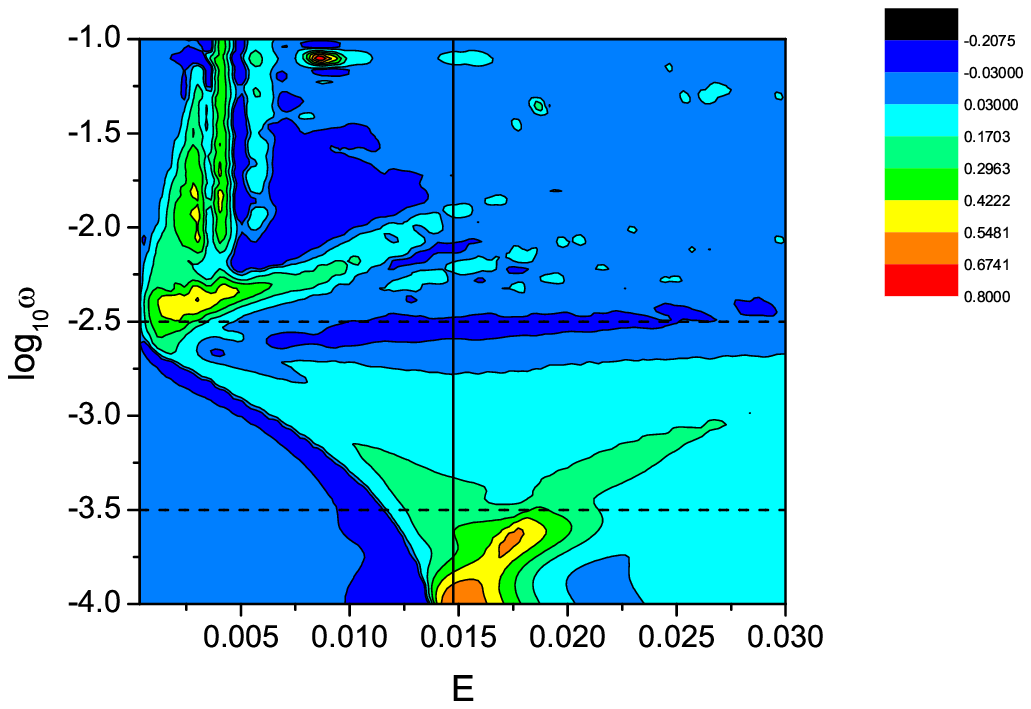}
\caption{\label{tcdiff} Contour plot of the difference between the classical and the quantum transmission coefficient $\Delta T = T_C - T_Q$ as a function of the energy $E$ of the incoming wave packet and of the frequency of the oscillation $\omega$. The height of the potential barrier is shown as a solid vertical line. The horizontal dashed lines correspond to frequencies $\log_{10}\omega=-2.5$ and $\log_{10}\omega=-3.5$ (see text).}
\end{center}
\end{figure}

In order to determine in more detail the region of frequencies $\omega$ and incident energies $E$ in which the agreement between classical and quantum mechanics is optimal, we perform a calculation of the classical $T_C$ and quantum $T_Q$ transmission coefficients in a two-dimensional grid of $E$ and $\omega$ values. We have used $60$ values for the frequency and $90$ values for the energy resulting in a $60\times 90$ grid on the $\omega-E$ plane. The results are shown in Fig.~\ref{tc2D}. We have also calculated the difference between the classical and the quantum transmission coefficient $\Delta T = T_C - T_Q$ shown in Fig.~\ref{tcdiff}.
From Fig.~\ref{tc2D} we observe that the classical and quantum transmission coefficients exhibit, at least qualitatively, a very similar behavior as a function of $E$ and $\omega$. A more quantitative description, given in Fig.~\ref{tcdiff}, shows transparently that there is good agreement between classical and quantum mechanics in a broad region of the $(E,\omega)$ plane while discrepancies occur in certain frequency regions. More specifically, in the low frequency region ($\log\omega < -3.5$), the discrepancy is enhanced at energies close to $V_0$ and is due to quantum tunneling and interference, whereas in the high frequency region  ($\log\omega >-2.5$), the discrepancy occurs at low energies. In this energy region resonances occur, and are due to the formation of a time-averaged potential having the form of a double barrier. In the intermediate frequency region ($-3.5<\log\omega<-2.5$) the agreement between classical and quantum mechanics is optimal almost in the whole energy range considered.

\begin{figure}
\begin{center}
\includegraphics[width=12cm]{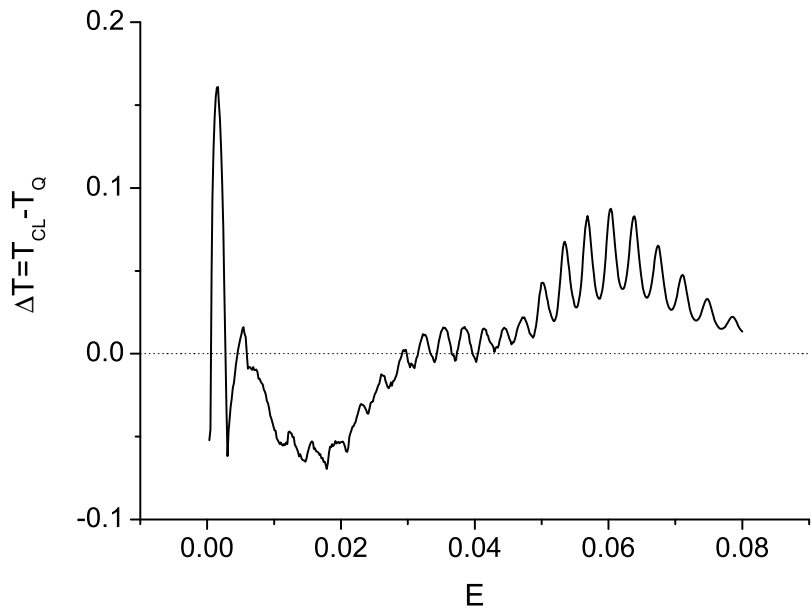}
\caption{\label{qc_diff_interm} The difference between the classical and quantum transmission coefficients $\Delta T = T_C - T_Q$ as a function of the energy $E$ of the incoming wave packet for $\omega=3\cdot 10^{-3}$.}
\end{center}
\end{figure}

Despite the overall optimal quantum-classical agreement in the intermediate frequency region, there are differences between the classical and quantum mechanics that can be attributed to kinematic effects. The difference between the classical and quantum transmission coefficients $\Delta T = T_C - T_Q$ as a function of the energy of the incoming wave packet is shown in Fig. \ref{qc_diff_interm}. In that plot we have extended the energy range to $0.08$ as at this value of the energy the classical transmission coefficient is $1$ and the quantum transmission coefficient approaches this value.
In the following we will give a brief kinematic interpretation of the differences between the classical and quantum transmission coefficients in the intermediate frequency regime. 

\begin{figure}
\begin{center}
\includegraphics[width=7.0cm]{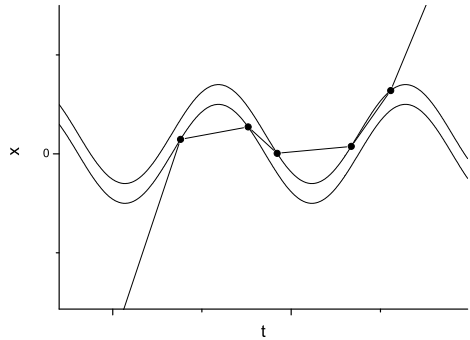}
\caption{\label{xt_interm} Schematic representation on the $x-t$ plane of an orbit exhibiting $5$ collisions with the oscillating barrier for  $\omega=3\cdot 10^{-3}$ and $E=0.02667$. The orbit is represented as joined linear segments and the boundaries of the barrier as sinusoidal curves. The points of impact are denoted with black dots.}
\end{center}
\end{figure}

\begin{figure}
\begin{center}
\includegraphics[width=9.5cm]{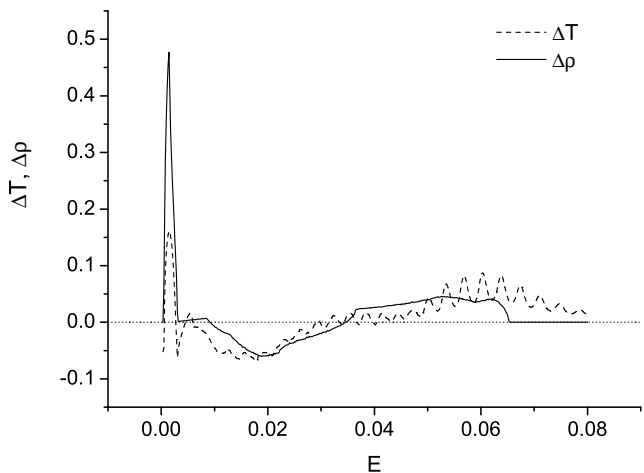}
\caption{\label{rho_qcdiff} $\Delta\rho=\rho _{ > 2,t}  - \rho _{ > 2,r}$ (solid line) and $\Delta T = T_C - T_Q$ (dashed line) as a function of the energy of the incoming wave packet for  $\omega=3\cdot 10^{-3}$.}
\end{center}
\end{figure}

In the intermediate frequency region, although there are no structures in phase space and as a consequence no dynamical trapping of the orbits, there is a significant fraction of orbits exhibiting more than two collisions. One such orbit is illustrated in the $x-t$ diagram of Fig. \ref{xt_interm}, where the orbit is represented by joined linear segments and the boundaries of the barrier as sinusoidal curves.
Qualitatively, due to interference effects, the presence of parts of the wave packet with many collisions that are finally transmitted , leads to a reduction of the quantum transmission coefficient compared to its classical value. Conversely, the presence of parts of the wave packet with many collisions that are finally reflected, leads to an enhancement of the quantum transmission coefficient compared to its classical value. For convenience let us denote as $\rho _{ > 2,t}$ and $\rho _{ > 2,r}$ the fraction of the total orbits with more than two collisions that are transmitted and reflected respectively. Following the above qualitative line of thinking, the difference $\Delta\rho=\rho _{ > 2,t}  - \rho _{ > 2,r}$ should be a measure of the difference between the classical and quantum transmission coefficient: increased $\rho _{ > 2,t}$ will enhance classical over quantum transmission whereas increased $\rho _{ > 2,r}$ will enhance quantum over classical transmission. This is indeed true, as can be seen from Fig. \ref{rho_qcdiff}, where $\Delta\rho$ is plotted on the same axes with $\Delta T = T_C - T_Q$. From that figure it can be seen that a purely kinematic measure, namely  the difference $\Delta\rho=\rho _{ > 2,t}  - \rho _{ > 2,r}$, describes quite accurately -except for the purely quantum resonance oscillations- the differences between the classical and quantum transmission coefficients.

\begin{figure}
\begin{center}
\includegraphics[width=9.5cm]{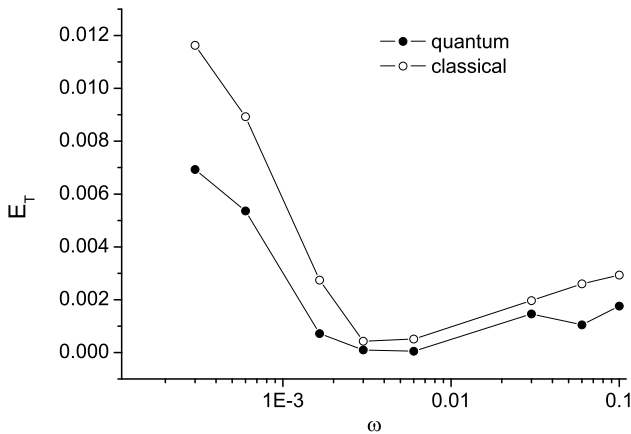}
\caption{\label{threshold} Classical (hollow circles) and quantum (full circles) transmission thresholds $E_T$ as a function of the oscillation frequency.}
\end{center}
\end{figure}

In the region of intermediate frequencies, apart from the enhancement of the agreement between classical and quantum mechanics, there is a significant lowering of the transmission threshold, i.e. the energy above which the transmission coefficient acquires a significant value. In order to illustrate this fact, we calculate the classical and quantum transmission threshold $E_T$, defined as the energy at which the transmission coefficient acquires for the first time the value $10^{-3}$. The results are shown in Fig.~\ref{threshold}. From this figure it becomes obvious that $E_T$ exhibits a minimum in both classical and quantum mechanics. Moreover, the frequency corresponding to this minimum ($\omega \simeq 3\cdot 10^{-3}$ in classical mechanics and $\omega \simeq 6\cdot 10^{-3}$ in quantum mechanics) is located in the intermediate frequency region, in which classical and quantum mechanics are in better quantitative agreement. 

\begin{figure}
\begin{center}
\includegraphics[width=7cm]{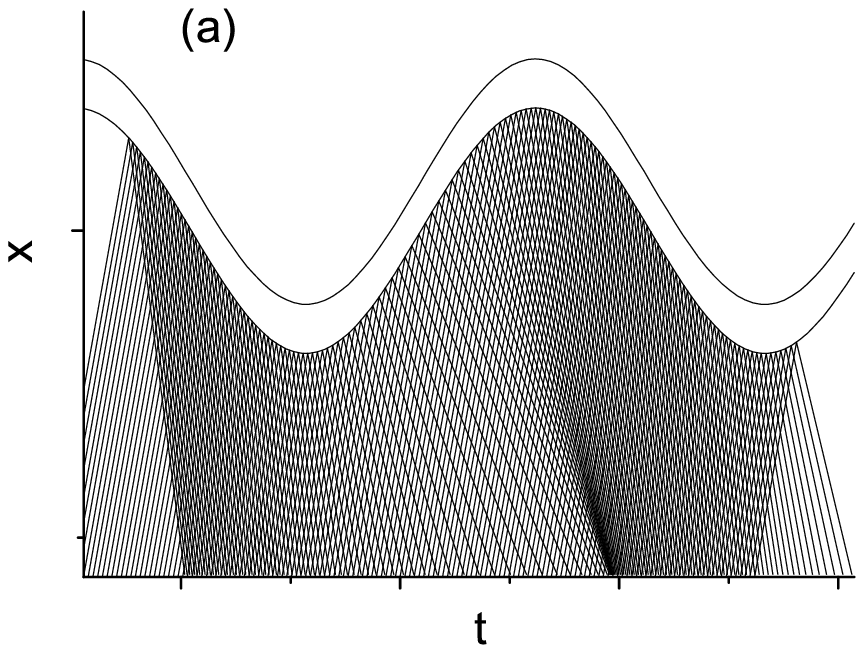}%
\includegraphics[width=7cm]{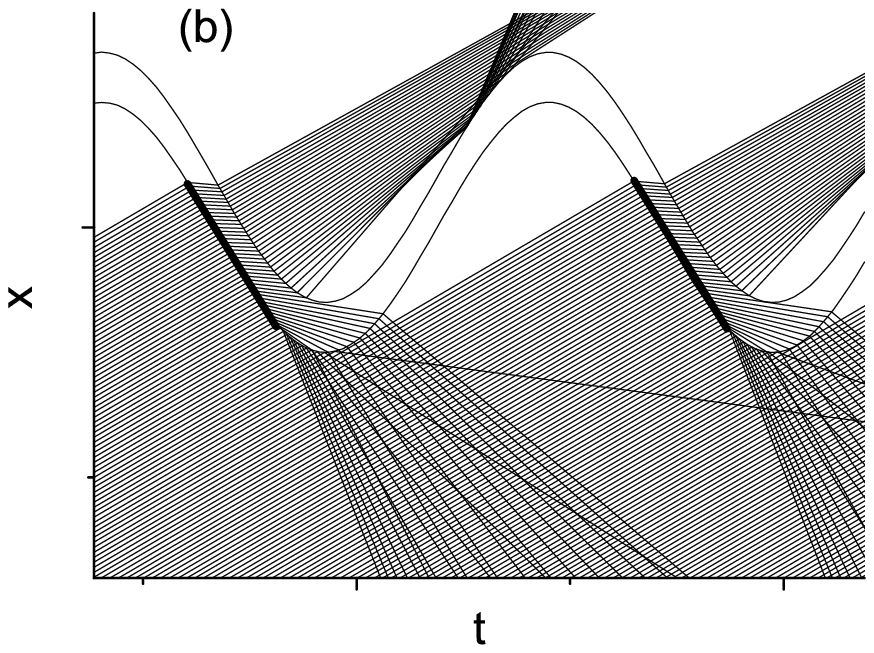}
\includegraphics[width=7cm]{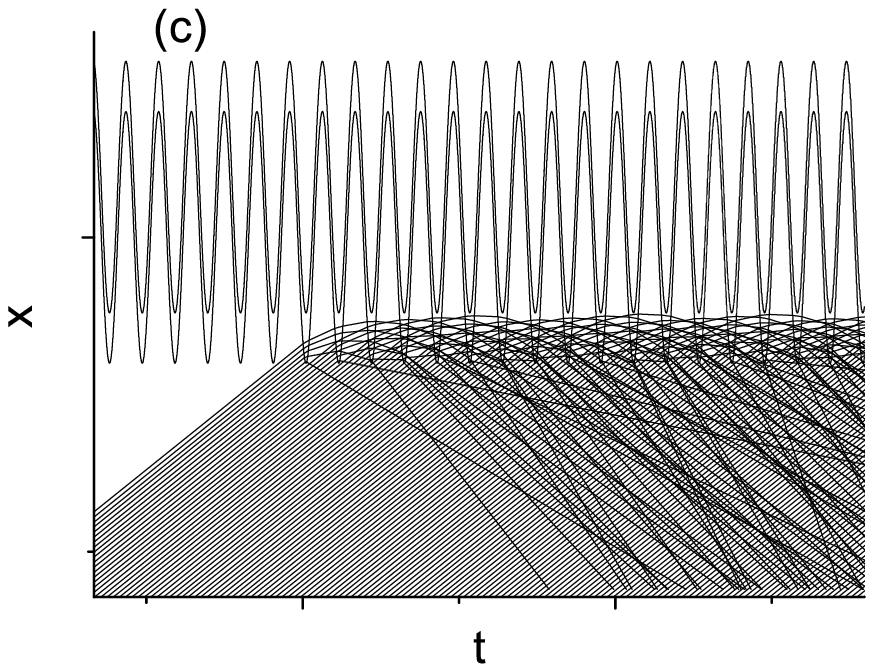}
\caption{\label{thresh_exp} Schematic representation on the $x-t$ plane of a part of a wave packet (bundle of initially straight lines) with $E=2\cdot 10^{-3}$ interacting with a laterally harmonically oscillating barrier (region between two sinusoidal curves) oscillating with frequency (a) $\omega=3\cdot 10^{-4}$, (b) $\omega=3\cdot 10^{-3}$ and (c) $\omega=3\cdot 10^{-2}$. In (b) the phases of the barrier for which transmission occurs are indicated with thick solid lines.} 
\end{center}
\end{figure}

The effect of the lowering of the transmission threshold as the frequency increases from the low to the intermediate frequency region admits a purely classical kinematic interpretation. In the range of intermediate frequencies ($\omega\simeq\omega_I$), particles with low energy can interact with the barrier at a phase close to $\pi$, leading to the observed small transmission threshold, since in this case the kinetic energy of the particle relative to the barrier is close to its maximal value.  This fact is illustrated in the $x-t$ diagram of Fig.~\ref{thresh_exp}(b), where the wave packet is represented as a bundle of initially straight lines and the boundaries of the barrier as sinusoidal curves. This figure corresponds to $\omega=3\cdot 10^{-3}$ and $E=2\cdot 10^{-3}$. The phases of the barrier for which transmission occurs are indicated with thick solid lines. At this point we should note that there are oscillation phases inaccessible to the scattered particle. This occurs when the velocity of the incoming particles is smaller than the maximum velocity of the barrier. The above effect that explains the reduced transmission threshold in the range of intermediate frequencies does not apply in the case of high and low frequencies. In the high frequency case, particles interact with the barrier at a phase close to $3\pi/2$, as shown in Fig.~\ref{thresh_exp}(c), and typically exhibit several collisions with the oscillating barrier. In the low frequency case, the corresponding $x-t$ diagram is shown in Fig.~\ref{thresh_exp}(a). In this case, all phases of the oscillating barrier are accessible to the scattered particle. However, at this value of the energy the transmission of particles is not allowed, since the maximum kinetic energy of the particles relative to the barrier is smaller than $V_0$.

Regarding the classical phase space of the system, it has been shown that in the high frequency region it exhibits mixed dynamics: it possesses a central island of stability centered on a periodic orbit, a structure of KAM islands and a thin layer of chaotic motion around them \cite{PAP05,KOC08}. It has been found that the central island appears for $\omega\geq  3.3\cdot 10^{-3}$ ($\log\omega\geq -2.48$). As it can be seen from Fig.~\ref{tcdiff} (upper left part of the graph), this frequency is very close to that above which classical - quantum disagreement occurs. Although the stable manifolds of the chaotic invariant set extend well outside the interaction region \cite{PAP05,KOC08}, the fraction of particles influenced by them is not statistically significant so as to induce structures (such as peaks or oscillations) in the classical transmission coefficient, unless a fine tuning in the initial conditions is made.

\section{Formation of coherent pulse trains}
\begin{figure}
\begin{center}
\includegraphics[width=7cm]{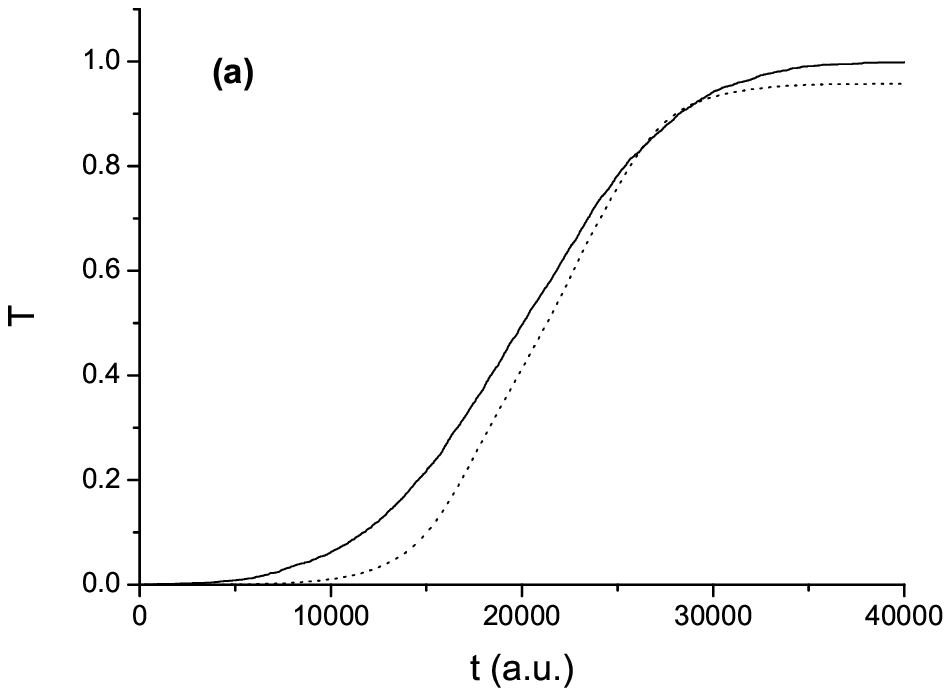}%
\includegraphics[width=7cm]{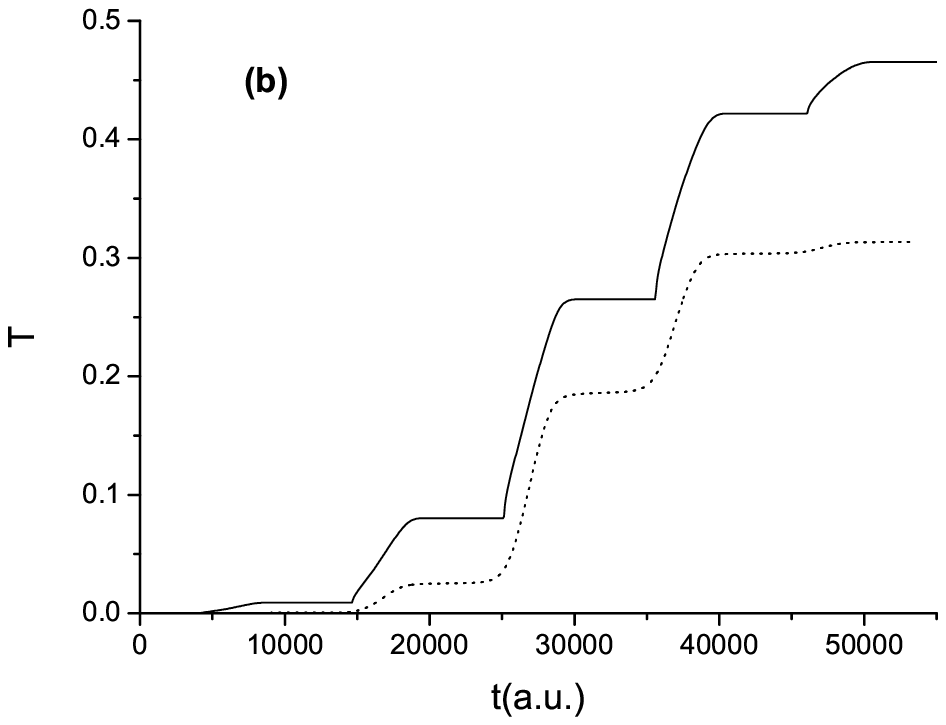}
\caption{\label{bar_tt} The classical (solid line) and quantum (dotted line) time-dependent transmission coefficient for $\omega=6\cdot 10^{-4}$ and (a) $E=0.03$ and (b) $E=0.013$.} 
\end{center}
\end{figure}

\begin{figure}
\begin{center}
\includegraphics[width=7.5cm]{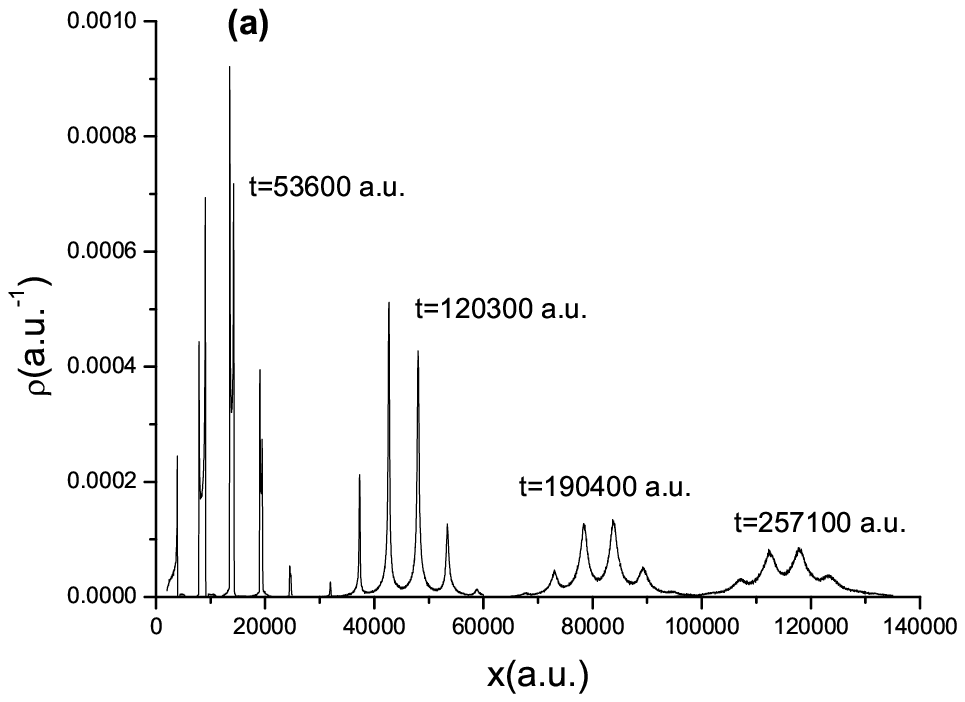}%
\includegraphics[width=7.5cm]{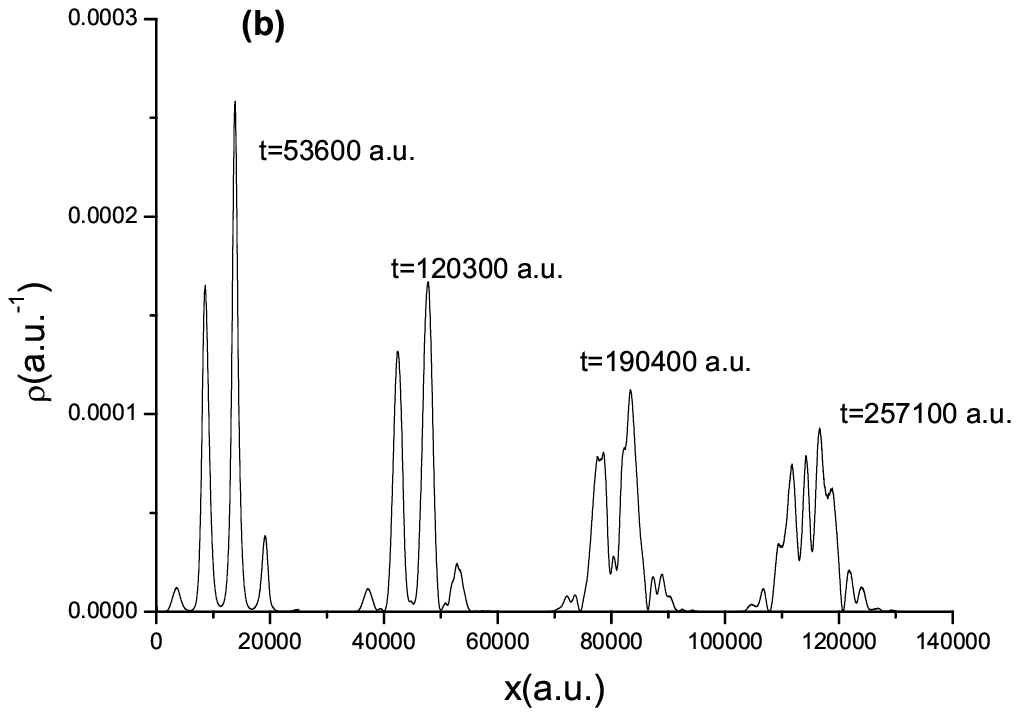}
\caption{\label{bar_snaps} Snapshots of the time evolution of the (a) classical and (b) quantum probability distribution of the transmitted wave packet for $\omega=6\cdot 10^{-4}$, $E=0.013$ and four values of time, namely $t=53600$, $t=120300$, $190400$ and $t=257100$.}
\end{center}
\end{figure}

In this section we will study QCC in an observable related to the space-time evolution of the scattered wave packet. More specifically, we focus our study on the form of the transmitted wave packet in space. The initial wave packet is the same as in the previous section and the frequency of the barrier oscillation is $\omega=6\cdot 10^{-4}$. In the following, we will study the time-dependent transmission coefficient, which is defined as the fraction of the initial wave packet that has been transmitted at time $t$:
\begin{equation}
T(t) = \int\limits_{A + \alpha }^\infty  {\left| {\Psi (x,t)} \right|^2 dx}.
\end{equation}
At the limit $t\to\infty$, $T(t)$ tends to the usual transmission coefficient. We have found that $T(t)$ can in general exhibit two distinct behaviors: it can either be a smoothly increasing function of $t$ or it can exhibit an interesting step-like structure. Two such representative cases are shown in Fig.~\ref{bar_tt} (a) and (b)  for incident wave packet energies $E=0.03$ and $E=0.013$ respectively. The step-like structure, on which we will focus in the following, appears at both classical and quantum levels.
In this case the transmitted wave packet splits to a series of well separated pulses. This is displayed in Fig.~\ref{bar_snaps}, which shows the time evolution of the classical and quantum probability distributions for the transmitted wave packet. For $t=53600$, the initial wave packet, after its interaction with the oscillating barrier, has been split into four narrower and well separated pulses. For $t=120300$ the pulses are still well separated, i.e. this splitting effect persists for time intervals much larger than the oscillation period ($T={2\pi\over\omega}\simeq 10472$) and for regions far from the boundaries of the interaction region. For longer times, the train of the four pulses loses its initial shape. This is mainly due to the fact that the pulses spread in position space due to the broadness of their momentum distribution (dispersion). Moreover, at the quantum level, quantum interference also contributes to the loss of the peaked structure of the pulse train. This can be seen in Fig.~\ref{bar_snaps} where for $t=257100$ the quantum probability density has almost lost its peaked structure whereas the classical probability density still retains four distinct peaks. Moreover, by performing a Fourier transform of the four pulses as they exit the interaction region, we find that their spectra are almost identical, i.e. the pulses are coherent.

\begin{figure}
\begin{center}
\includegraphics[width=6.5cm]{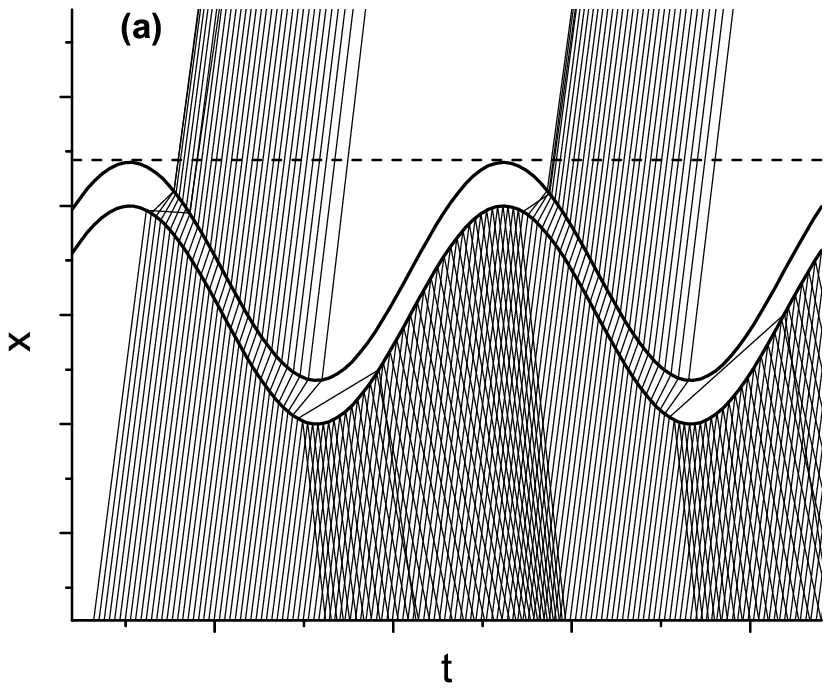}%
\includegraphics[width=6.5cm,,bb= 3 160 517 562]{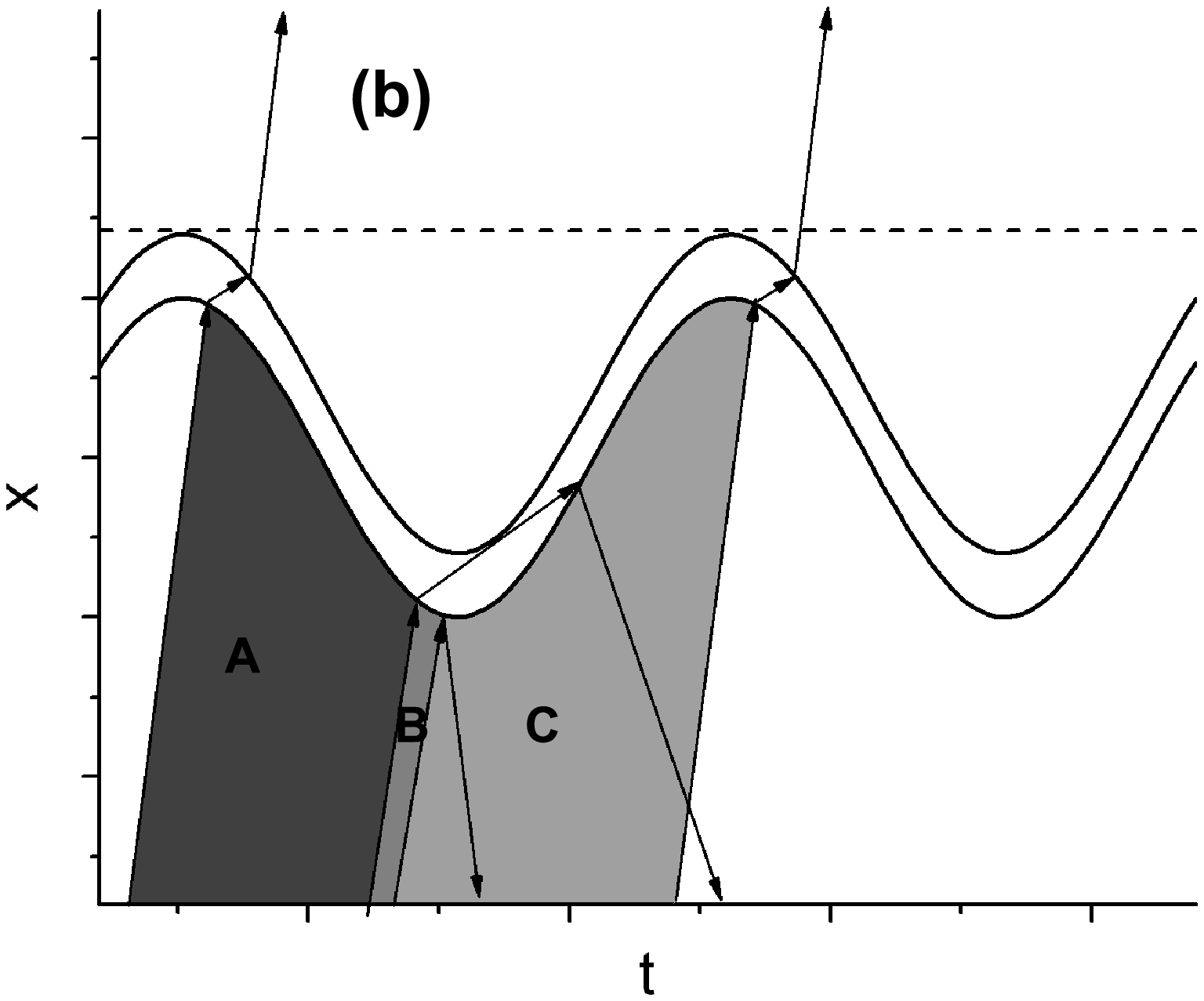}
\caption{\label{bar_interp} (a) Schematic representation on the $x-t$ plane of a part of a wave packet (bundle of initially straight lines) interacting with a laterally harmonically oscillating barrier (region between two sinusoidal curves). The right boundary of the interaction region is shown with a dashed line. (b) Schematic representation of the regions A, B, C: the orbits in region A are transmitted whereas the orbits in regions B and C are reflected (see text).}
\end{center}
\end{figure}

The splitting of the incident wave packet into distinct pulses admits a purely classical kinematic interpretation. In the parameter region where the splitting occurs, three distinct dynamical behaviors for the orbits of the particles can occur depending on the phase of the barrier at the instant of the collision. In Fig.~\ref{bar_interp}(a) a representation of the wave packet motion in a $x-t$ diagram is shown. The regions marked as A,B and C in Fig.~\ref{bar_interp}(b) correspond to the three distinct dynamical behaviors. The orbits of region C do not have enough kinetic energy (relative to the barrier) to overcome its height and therefore they are reflected. For these orbits
\begin{equation}
\left( {v_0  - v_b } \right)^2  < \frac{{2V_0 }}{m},
\end{equation}
where $v_0  = \sqrt {2E/m}$ is the velocity of the particle before the collision and $v_b$ is the velocity of the barrier at the instant of the collision. In contrary to the orbits of region C, for the orbits of regions A and B the condition
\begin{equation}
\left( {v_0  - v_b } \right)^2  > \frac{{2V_0 }}{m}
\end{equation}
is fulfilled and the particles enter the barrier. However, some of these orbits (region B), after their collision with the barrier, do not have enough energy to reach the right boundary of the barrier and they collide for a second time with the left boundary. These orbits are finally reflected as well. There is therefore a time interval during which no particles reach the right boundary of the interaction region ($x=A+\alpha$), shown with a dashed line in Fig.~\ref{bar_interp}(a) and (b). A single pulse in the transmitted wave packet corresponds to a period of the oscillation and therefore contains a succession of regions A, B, C in the incident wave packet. The wider the wave packet in position space is, the more pulses appear after its interaction with the oscillating barrier.

The existence of the regions A, B, C and the occurrence of the pulse splitting effect depend on the parameters of the system. In order for $T(t)$ to exhibit steps and, as a consequence, the incoming wave packet to be split in several pulses, the incoming wave packet has to probe all the phases of the oscillating barrier. This requirement introduces the constraints
\begin{equation}
\sigma _x  > v_0 \tau 
\end{equation}
and
\begin{equation}
v_0  > A\omega,
\end{equation}
where $\tau$ is the period of the oscillation. Moreover, during a period of the oscillation, both reflection and transmission of particles should occur. The latter introduces the additional constraint for the initial velocity:
\begin{equation}
\sqrt {\frac{{2V_0 }}{m}}  - A\omega  < v_0  < \sqrt {\frac{{2V_0 }}{m}}  + A\omega .
\end{equation}
For smaller velocities all orbits are reflected whereas for larger velocities all orbits are transmitted. As for the number of steps, and therefore the number of pulses, it is found that it does not change by varying $E$ and $V_0$ but it can be adjusted by varying the frequency of the oscillation: with increasing frequency, the number of steps increases and therefore their width decreases, i.e. the number of transmitted pulses increases and their width in space decreases.

We conclude that it is possible to prepare the desired pulse-splitting effect for a given system with parameter values $(V_0,m,\alpha)$ by matching the parameters $(A,\omega)$ of the external driving field as well as the initial spread $\sigma_x$ and energy $E$ of the incoming wave packet using a purely classical calculation. One possible experimental setup for the observation of the pulse splitting effect are semiconductor heterobarriers which are driven either by an external laser field or an applied AC gate voltage \cite{DIT98,SUN98}. The wave packet of quasiparticles could be created by a (second) laser allowing to control the width and energy of the initial pulse. Observation of the transmitted time-delayed pulses can be done via time-resolved measurements. A potential application would be the use of the pulse splitting effect to build controllable intermittent pulse sources on a nanometer scale which should be of interest to nanoelectronics. Due to the very general character of the scattering process off oscillating barriers, the mechanism for the formation of pulse trains might occur in a variety of other physical systems of either classical or quantum character, e.g. for (cold) atoms encountering oscillating barriers or penetrable walls.

\section{Conclusions}
In this work we have studied the scattering of classical and quantum wave packets off a 1-dimensional barrier laterally oscillating with a harmonic time law. Our study has been mainly focused on Quantum Classical Correspondence (QCC) and its dependence on the parameters of the system. More specifically, we have considered the transmission coefficient as well as the form of the transmitted wave packet in both classical and quantum mechanics. 

Regarding the transmission coefficient, the region of parameter space where classical and quantum mechanics are in good agreement has been investigated. It is found that in a certain frequency region (the intermediate one), QCC is optimal almost in the whole energy range considered and the transmission threshold exhibits a minimum. The latter admits a purely classical interpretation based on kinematic effects. Moreover, in the same frequency range, the difference between the classical and quantum transmission coefficient can be described mainly by kinematic arguments as well.

Regarding the form of the transmitted wave packets, it is found in both classical and quantum mechanics that in a rather broad region of parameter space, the incoming wave packet can be split into a train of well separated coherent pulses. This effect is not related to the phase space of the underlying classical system and admits as well a purely classical kinematic interpretation. The pulse splitting effect can possibly be observed experimentally, for example in appropriately driven semiconductor heterostructures, and is expected to have useful applications.

\end{document}